\documentclass{llncs}
\usepackage{graphicx}
\usepackage{xkeyval} 
\usepackage{svg}
\usepackage{float}
\usepackage{hyperref}
\usepackage{cite}
\usepackage{soul}
\graphicspath{ {./figures/} }

\makeatletter
\define@key{PAX@Gin}{scale}{}
\makeatother

\newcommand{\ignore}[1]{}

\begin{document}
  \title{Analyzing CNN Based Behavioural Malware Detection Techniques on Cloud IaaS}
  \author{Andrew McDole\inst{1}  \and Mahmoud Abdelsalam\inst{2} \and Maanak Gupta\inst{1} \and Sudip Mittal\inst{3}}
  \institute{Tennessee Technological University, Cookeville, TN, USA
  \and Manhattan College, Riverdale, NY, USA \and University of North Carolina at Wilmington, NC, USA
  \\\email{Email: amcdole42@students.tntech.edu, mabdelsalam01@manhattan.edu, mgupta@tntech.edu, mittals@uncw.edu}}
  \maketitle
  
  \begin{abstract}
      Cloud Infrastructure as a Service (IaaS) is vulnerable to malware due to its exposure to external adversaries, making it a lucrative attack vector for malicious actors. A datacenter infected with malware can cause data loss and/or major disruptions to service for its users. This paper analyzes and compares various Convolutional Neural Networks (CNNs) for online detection of malware in cloud IaaS. The detection is performed based on behavioural data using process level performance metrics including cpu usage, memory usage, disk usage etc. We have used the state of the art DenseNets and ResNets in effectively detecting malware in online cloud system. CNN are designed to extract features from data gathered from a live malware running on a real cloud environment. Experiments are performed on OpenStack (a cloud IaaS software) testbed designed to replicate a typical 3-tier web architecture. Comparative analysis is performed for different metrics for different CNN models used in this research.
  \end{abstract}
  
  \section{Introduction and Motivation}
  Cloud has become a popular platform due to its characteristics of on-demand services, infinite resources, ubiquitous availability and pay-as-you go business model \cite{mell2011nist}. Infrastructure as a Service (IaaS) is the most widely offered service model where the resources of a large data center can be purchased by clients to perform computing tasks. 
  Since user clients can utilize any number of virtual machines, ranging from a couple to thousands, automatic monitoring of these virtual machines is necessary to ensure the security of the cloud provider and its clients. While there are several risks associated with IaaS, one of the greatest risks is the possibility of a virtual machine becoming infected with malware and spreading the malware to other virtual machines in the data center. This would put cloud providers and their customers in danger as well as end users whose data is stored or transferred on these infected virtual machines. As cloud providers increase their client base, the potential for loss also increases and so does the responsibility of cloud providers to invest in security mechanisms for their customers. The scale of an attack is multiplied due to similar configuration and automatic provisioning of the virtual machines (VMs) hosted by a cloud service provider. Identical configurations for these virtual machines make attacks repeatable and allow them to more likely spread within the data center once a single machine is infected.
  \\\indent Static malware analysis technique is widely used, in which the files are scanned before they can be executed on the systems. In such case, file is disassembled by a disasssemblers to obtain the source code which can then be examined using different tools. Although the method is fast and efficient, but it can be easily dodged by malware writers who can trick the disassemblers into generating incorrect code. This is done by inserting errors which lead to the actual code execution path being hidden or obfuscated. The binary file can also be worked on directly. An example of this is extracting n-grams of the binary file as features and then using machine learning techniques to locate known malicious patterns. Static analysis generally fails in the case of cloud malware as malware is injected into an application that was already scanned and deemed safe. Such an attack in cloud IaaS is referred to as a cloud malware injection \cite{gruschka2010attack}. In this case, if the application is not re-scanned at a later time, the newly inject malware will not be detected. Therefore, the need to constantly monitor these applications running in cloud environments is essential.
  \\\indent While there are several works in the domain of malware detection, few research papers \cite{abdelsalam2018malware,abdelsalam2017clustering,abdelsalam2019online,pannu2012aad,dawson2018phase,wang2009ebat,watson2015malware} deal with \emph{online} malware detection specifically and  in particular provide solutions using machine learning based approach. This process consists of a typical machine learning approach i.e. building a machine learning model, training the model with relevant dataset captured, and using the trained model to determine if a malware exists in the system or not. In the case of building the model, features must first be selected to determine what data will be used as input. This is no different for cloud based detection methods except that the features to be chosen are limited to the information that can be gained through the hypervisor. Through careful selection of features, machine learning can be used to provide dynamic malware analysis and detect in case machines have been infected by adversaries in the data centers. This kind of dynamic analysis fulfills the need for constant surveillance in cloud IaaS for malware detection.
  \\\indent The most unique characteristics of cloud computing include  resource pooling, on-demand self-service, and rapid elasticity which can be fulfilled by an auto-scaling architecture. In this paper, we focus on auto-scaling wherein the machines are spawned based on the demand and usually these VMs are of similar type, resulting in similar behaviour. It is likely that that an injected malware will result in behaviour deviation
  on a VM at some point. In this work, we seek to detect such malicious behaviour and compare state-of-the-art deep learning models on several parameters. We are focused on detecting only one VMs which have been compromised ignoring that the fact that all similar VMs can be infected by an adversary in a more sophisticated attack. We plan to work on this as a next step to this problem.
 
  This work is an extension to our earlier work where only one kind of CNN model was used, with the prime goal that such techniques can be effectively used malware analysis. In this work, we compare and contrast several CNN models using the same data as \cite{abdelsalam2017clustering,abdelsalam2018malware,abdelsalam2019online} and six other deep learning models to determine possible use cases within a cloud IaaS scenario. For all models, the dataset consists of process-level metrics collected from the virtual machine hypervisor. Since these models are CNNs, the data is formatted as two dimensional matrices with the dimensions being $unique\: processes \times selected\: features$. Since many of our models require the input to be 3 dimensional shape, the 2d matrix is copied to fulfill the third dimension requirement.

  
  \indent The paper is organized as follows. Section \ref{Related Work} discusses related work in cloud online malware detection. Section \ref{Key Intuition and Methodology} provide an overview of the key intuition and methodology for the experiments. Section \ref{Qualitative Analysis and Comparison} covers evaluation metrics and experimental results whereas Section \ref{Discussion} presents comparative analysis among different CNN models used. Section \ref{Limitations} covers certain limitations of our approach along with discussion on future work. Finally, section \ref{Conclusion} summarizes this paper.
  
  \section{Related Work} \label{Related Work}
  Several works have been done in malware detection which focus on different aspects of several approaches. The first step in  developing a machine learning based model for online malware detection is to  determine which features are most relevant and are to be extracted. Research papers  \cite{10.5555/2483628.2483648,pirscoveanu2015analysis,tobiyama2016malware} focus on API calls whereas \cite{7942417,10.1007/978-3-642-33704-8_21,dawson2018phase} primarily utilize system calls. Other features such as performance counters \cite{demme2013feasibility} or memory features \cite{khasawneh2015ensemble,xu2017malware} have also been used. Although several existing resilience frameworks exist \cite{sterbenz2010resilience,watson2013towards,watson2015malware,marnerides2015multi}, it is likely that novel attacks and new techniques will defeat existing detection methods.
  \\\indent Most of the algorithms for detecting malware, such as support vector machines (SVM)\cite{watson2015malware}, all-nearest-neighbor (ANN) classifier\cite{fan2016malicious}, and na\"ive bayes\cite{firdausi2010analysis,10.5555/2483628.2483648}, work for examining a single VM in the cloud. While a running single vm is not the expected use case of cloud environments, there is virtually no difference between a single VM and a standalone host when it comes to detecting malware on them. Generally, most works\cite{7942417,10.5555/2483628.2483648,pirscoveanu2015analysis,tobiyama2016malware,10.1007/978-3-642-33704-8_21,dawson2018phase,demme2013feasibility,khasawneh2015ensemble,xu2017malware} focus on features that can be extracted through the hypervisor.
  Dawson et al \cite{dawson2018phase} collect system calls for features and are primarily concerned with rootkits. A non linear phase-space algorithm is used in their analysis of system calls to detect anomalies. The results are evaluated on the phase-space graph dissimilarities.
  \\\indent Entropy based Anomaly Testing (EbAT) was introduced in  \cite{wang2009ebat}. EbAT analyzed multiple metrics such as CPU and memory utilization for the purposes of anomaly detection. The paper analyzed these metrics based upon distribution instead of a flat threshold. This approach yielded accurate results for detection and the ability to scale to keep up with metric processing. However, the evaluation did not demonstrate usefulness in practical and realistic cloud environment scenarios. Azmandian et al. \cite{azmandian2011virtual} utilize performance metrics such as disk and network input-output gathered from the hypervisor to form a new anomaly detection approach. K-NN and Local Outlier Factor are unsupervised machine learning techniques used in this work.
  
  \indent Work by Abdelsalam et al \cite{abdelsalam2017clustering} showed that a black box approach can be used to detect malware. This paper used VM-level performance and resource utilization metrics. This approach worked well in detecting highly active malware which showed up in the resource utilization metrics, but was not as effective in detecting malware that hid itself with low utilization. Similarly, in \cite{abdelsalam2018malware} the authors introduce a detection method which uses a CNN model with the goal of identifying low profile malware. This method achieved ~90\% accuracy using resource metrics and was able to identify multiple low-profile malware. While these results are good, it is limited in that it targeted only a single virtual machine like many other related works without features like auto-scaling.
  \section{Key Intuition and Methodology}\label{Key Intuition and Methodology}
  In this section, we discuss the key intuition behind our approach and describe our methodology in detail. 
  \subsection{Key Intuition}
  So as to detect online malware using process-level information, we train a model on a dataset that contains benign and malicious samples. Each sample consists of information about a process or collection of processes, and the task is to classify the input sample as benign or malicious. To build up our dataset of benign samples, we run a Virtual Machine (VM) normally without the presence of malicious software. Malicious data samples are collected after the VM has been infected with malware. 
  
  Different malware are used for different runs of the experiment to create the dataset. We then partition our dataset into into training, validation, and testing datasets. In other words, the model is trained on samples from different experiments which contained different malware. This way, the model generalizes itself to detect different malware through the various ways they reveal themselves in process metrics. A model's ability to generalize and predict new samples is dependent on its internal architecture. More complex models may achieve higher accuracy by adding more hidden layers or by connecting those hidden layers in a novel manner. 

  \subsection{Methodology}
\begin{figure}[t]
  \centering
    \begin{minipage}{.5\linewidth}
      \centering
      \includegraphics[width=.6\linewidth]{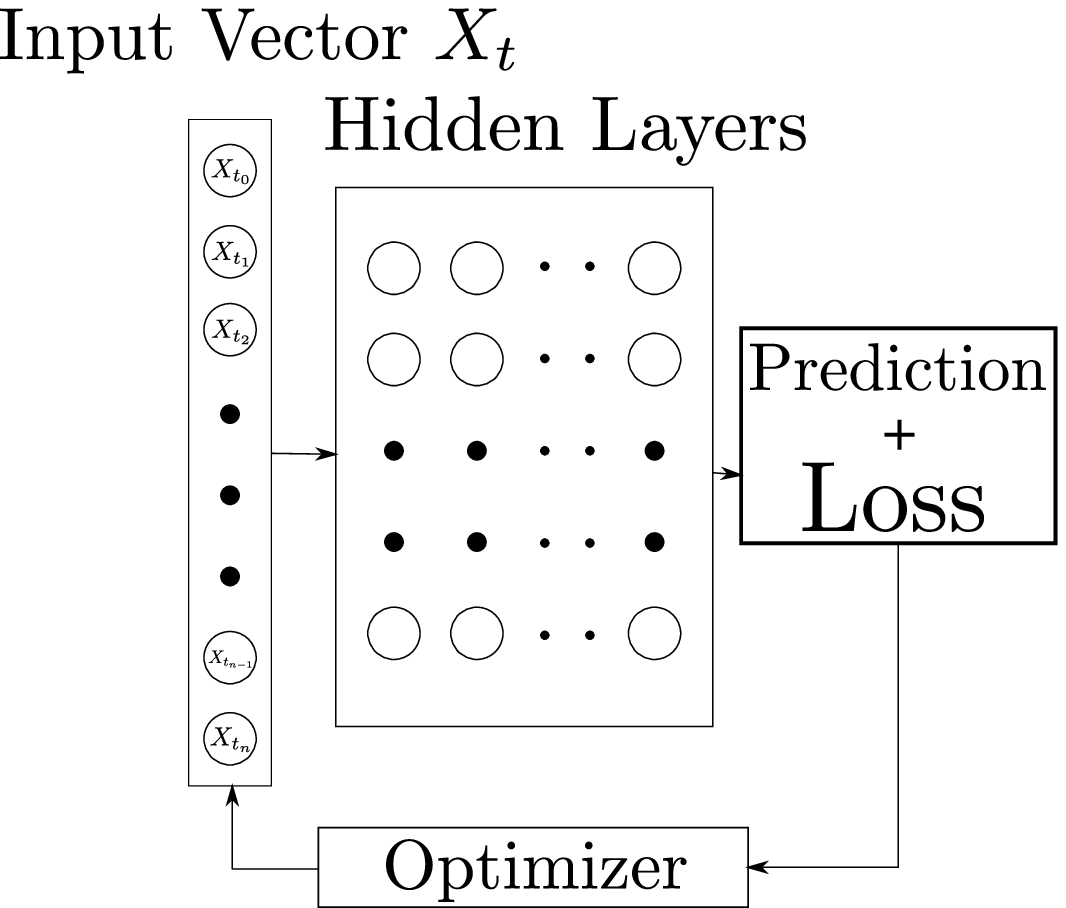}
      \caption{Neural Network Flow}
      \label{fig:standard_model_input}
    \end{minipage}%
    \begin{minipage}{.5\linewidth}
      \centering   
      \includegraphics[width=.75\linewidth]{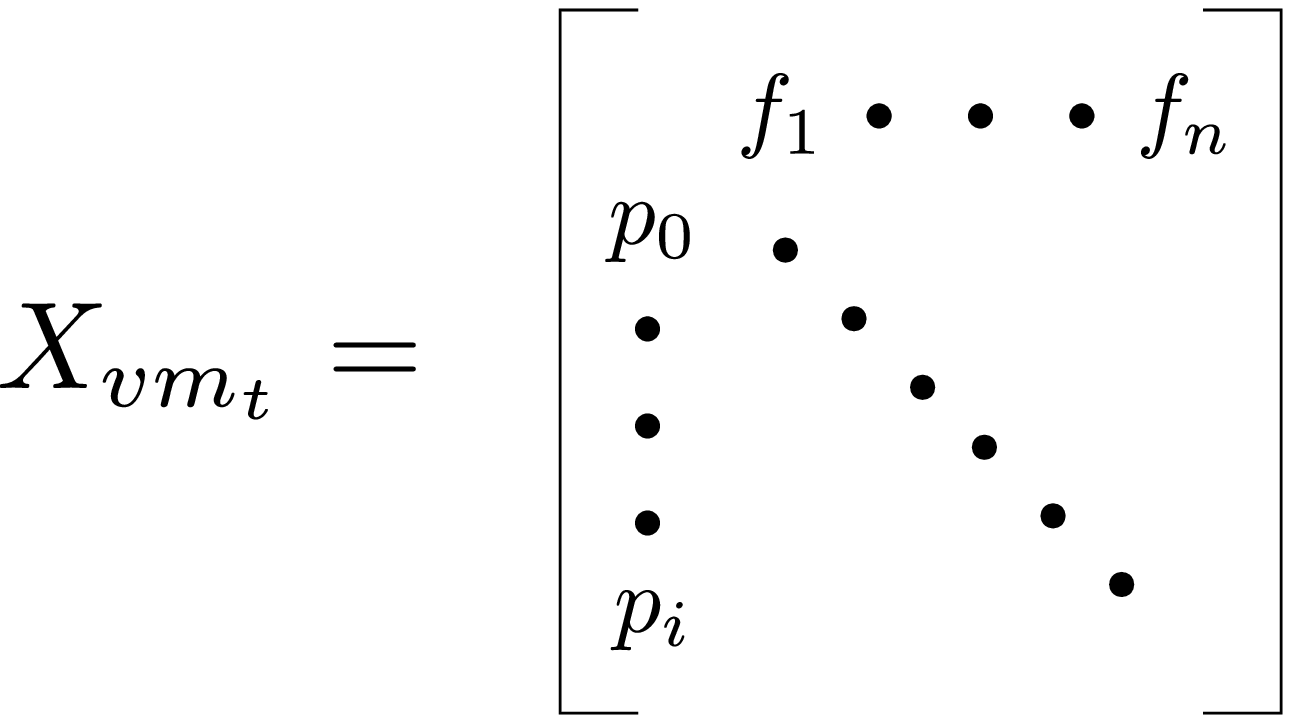}
      \caption{Sample at Time $t$ Consisting of Process Level Information}
      \label{fig:process_level_diagram}
    \end{minipage}
  \end{figure}

  Convolutional Neural Networks (CNN) have been commonly used in various visual imagery tasks. A basic flowchart of a neural network is shown in \autoref{fig:standard_model_input}. CNNs generally take two dimensional data as an input, in our case, the process level data is represented as a two dimensional array. A sample consists of rows of processes with columns of process features. Assuming $p_i$ is a process, $f_n$ is a process metric, $vm$ is a virtual machine ID, then $X_{vm_t}$ is a sample at time $t$ as shown in \autoref{fig:process_level_diagram}.
  
  Each sample represents a single virtual machine at a given time interval so the models learn what an infected machine ``looks like'' over time. During the course of time in an operating system, processes get created and destroyed and as these IDs can be assigned/re-assigned to different processes, they provide no useful information for the task at hand. For this reason, we focus on  \emph{unique process} defined as a tuple that contains a process ID, the command used to run the process, and a hash of the binary executable. This \emph{unique process} will be referred to as a process in this work. 
  Once the training dataset has been used to train the model, it is used for generating predictions on an unseen test data set that the model did not use during the training process.
  

  We used Openstack\footnote{Openstack. \url{https://www.openstack.org/}}, a popular cloud computing platform to replicate a standard 3-tier web architecture consisting of a web server, application server, and a database. Auto-scaling was enabled on the web server and the application servers were configured with a policy based on the average CPU utilization of the VMs. As per the policy, if the average CPU utilization is above 70\%, the architecture scales out and it scales in if the utilization is below 30\%. We spawned between 2 and 10 servers in each tier depending on the traffic load. An ON/OFF Pareto distribution with the default NS2\footnote{NS2 Manual. http://www.isi.edu/nsnam/ns/doc/node509.html.} tool parameters was used to generate the traffic load.
  
\begin{figure}[t]
      \centering
      \includegraphics[scale=0.6]{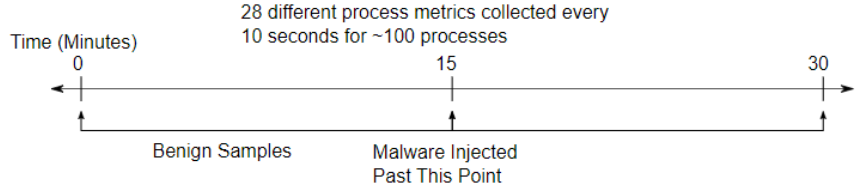}
      \caption{Data Collection Overview}
      \label{fig:data_collection}
  \end{figure}

  \autoref{fig:data_collection}, shows the data collection process. Each experiment was 1 hour long, consisting of a 30 minute clean phase and a 30 minute infected phase. During the clean phase, the virtual machines were untouched. During the infected phase, malware was injected into a virtual machine at some time after the infected phase started. We introduced 113 different malwares to collect our dataset. These malwares were obtained from VirusTotal\footnote{VirusTotal Website. https://www.virustotal.com.}. The VMs were configured with full internet access and all firewalls were disabled. This was done so that the malware could operate without any interference. After every 10 seconds, a sample was collected from the infected virtual machine in the experiment resulting in 360 samples over the course of each experiment.
  
  \begin{figure}[t]
    \centering
    \includegraphics[scale=0.5]{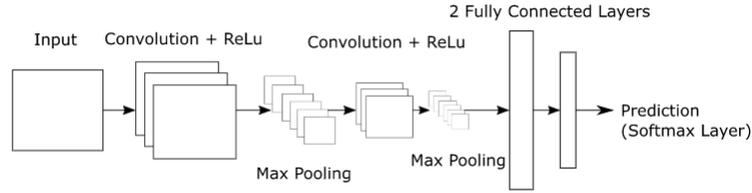}
    \caption[]{LeNet-5 Model}
    \label{fig:lenet5}
  \end{figure}
  \subsection{Convolutional Neural Network Models}
  \subsubsection{LeNet-5 \cite{lecun1998gradient}:} It is an example of a shallow CNN. It has few layers so the gradients can be computed quickly. \autoref{fig:lenet5}, shows the model architecture. Note that the architecture is simple and straightforward where the output of each layer serves as the input to the next layer.
  
  The input to the model, would be a 2 dimensional matrix of 120x45 representing a sample with a maximum of 120 processes and 45 features of these processes. Each process that was not active at the time the sample was taken, but would become active during the course of the experiment was padded with zeroes. The first layer of LeNet-5 consists of a convolutional layer with 32 kernels, each with a size of 5x5. The output of this layer is 32 feature maps with the same input shape of 120x45. The max pooling layer of size 2x2 downsizes these feature maps to become 60x23. The second convolutional layer has 64 kernels with the shape of the output from the previous max pooling layer, 60x23. This convolutional layer is followed by another max pooling layer of size 2x2 which results in 64 feature maps of size 30x12. The final layers of LeNet-5 are fully connected with sizes 1024, 512, and 2 respectively. The final layer has an output of size 2 since it represents a binary prediction of malicious or benign sample.
  
All of the activation functions used Rectified Linear Units (ReLU) \cite{agarap2018deep} which were placed after every convolution and fully connected layer excluding the final layer. We used the Adam Optimizer \cite{kingma2014adam}, which is a stochastic gradient descent algorithm with automatic learning rate adaptation. This optimizer trains the weights of the model after every min-batch. The learning rate controls how drastically the weights of the model are changed in response to the backpropagation. A higher learning rate leads to faster training but can result in unstable gradient descent and can inhibit convergence. A learning rate that is too slow can cause the model not to achieve higher accuracy results.

    \begin{figure}[t]
      \centering
      \begin{minipage}{.5\linewidth}
      \centering
      \includegraphics[scale=0.2]{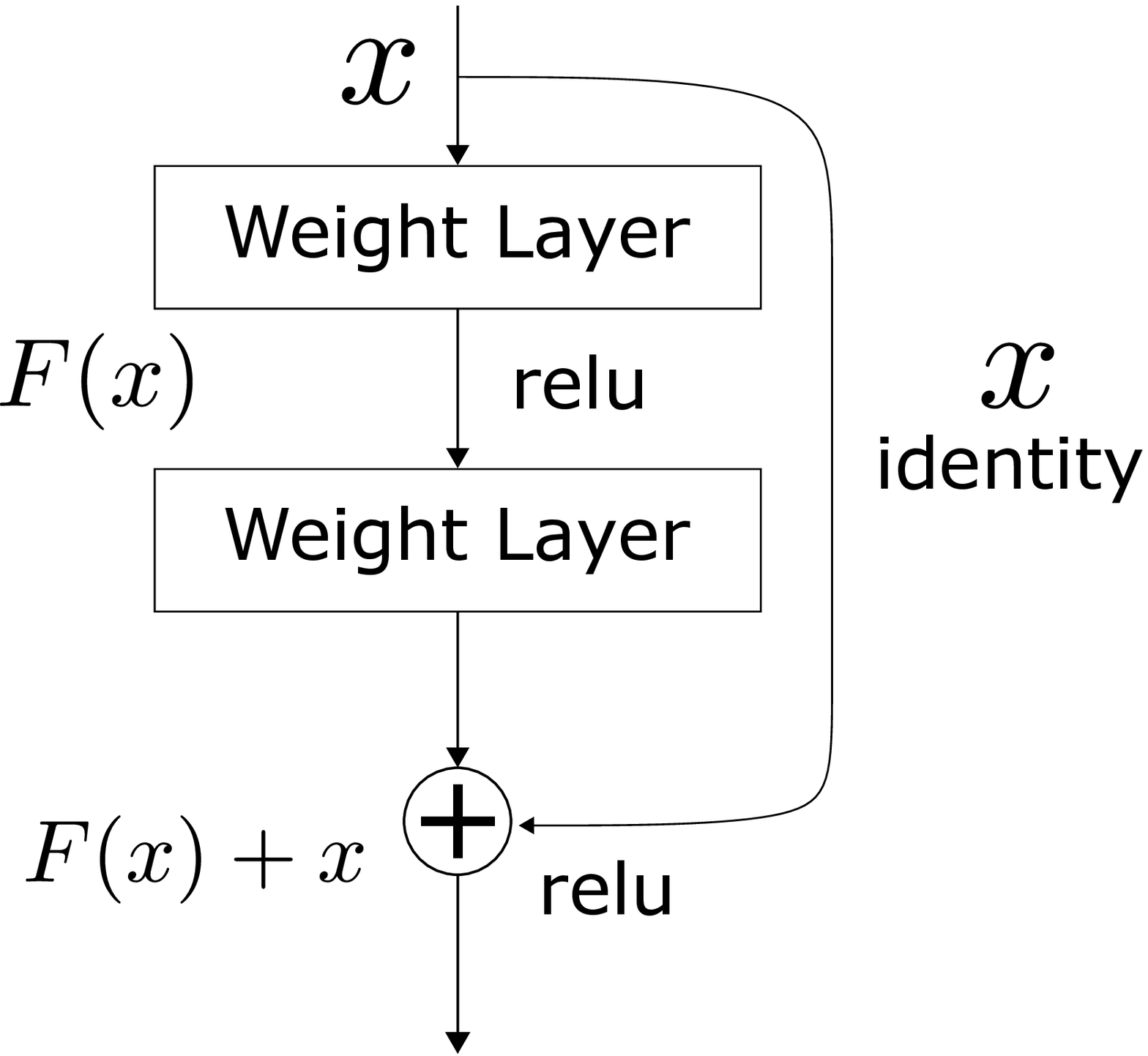}
      \caption{Residual Block Diagram}
      \label{fig:residual_block_diagram}
    \end{minipage}%
    \begin{minipage}{.5\linewidth}
      \centering   
      \includegraphics[scale=0.6]{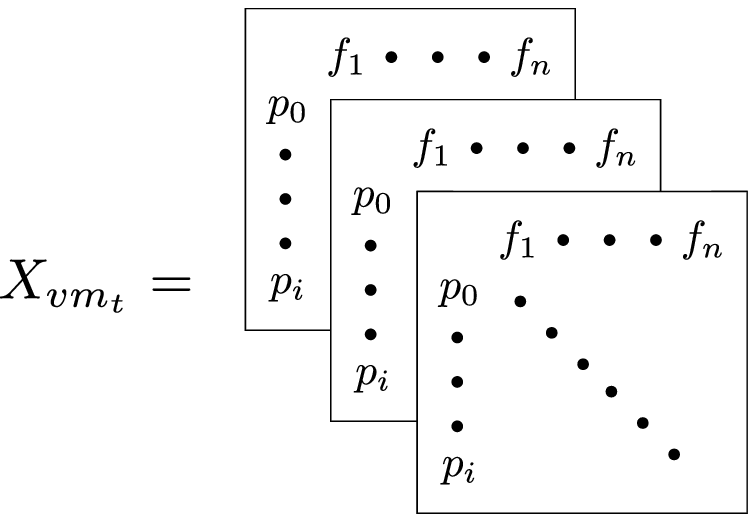}
      \caption{Data Input Shape with Window Size 3}
      \label{fig:data_shape_3d}
    \end{minipage}
  \end{figure}
  \subsubsection{Residual Networks:}
  One problem with models with a large number of layers is degradation \cite{He2015}. This is observation that adding more layers to the network can lead to optimization problems and therefore lower accuracy. This degradation is caused by the backpropagation not being able to reach the initial layers of the model. Residual networks (ResNets) solve this issue by adding skip-connections or residual connections. By adding these shortcut paths between layers, the gradient is allowed to flow better through the model and deeper models are able to be trained without degradation.

Residual blocks as shown in \autoref{fig:residual_block_diagram}, are used in \textit{ResNets}\cite{He2015}. The \textit{identity} is the shortcut connection and it is what allows the back propagation to affect the initial layers and allow them to learn as quickly as the final layers in the model. Three ResNets were used in our work: ResNet-50, ResNet-101, and ResNet-121. Each ResNet required the window size of the data to be three, but the samples were all 2d matrices. All samples had their data replicated twice more to form 3 dimensional data. A representation of this data is shown in \autoref{fig:data_shape_3d}. At the end of each model, global average pooling was added.
  \begin{figure}[t]
    \centering
    \includegraphics[scale=0.2]{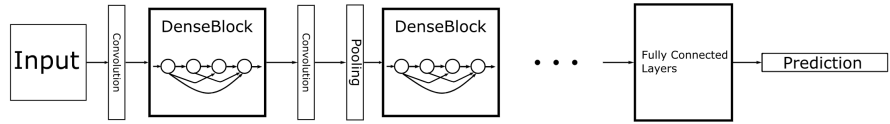}
    \caption{Dense Networks}
    \label{fig:densenet_diagram}
  \end{figure}
  \subsubsection{Dense Networks:}
  Where ResNets seek to resolve the gradient degradation problem, DenseNets \cite{DBLP:journals/corr/HuangLW16a} attempt to alleviate the vanishing gradient problem\cite{DBLP:journals/corr/abs-1211-5063}. A generic DenseNet model is shown in \autoref{fig:densenet_diagram}. DenseNets are different from ResNets because instead of having an identity mapping from one layer to the next, DenseNets pass the outputs of each layer to all subsequent layers. This way, each layer has collective knowledge from all the preceding layers. This causes the feature maps to be `reused' by latter layers. Due to this reuse of feature maps, less feature maps are required as input due to the compounding nature of DenseNets.
  \\\indent Each dense block makes use of these identity mappings and feature reuse. Between each dense block, there are transition layers that are comprised of a convolution and pooling layer. These are meant to reduce the feature map size between dense blocks. Similar to the ResNets, all DenseNets models received the same input shape as the ResNet models, 120 $\times$ 45 $\times$ 3. The batch size used was 64 for all models to maintain consistency.
  \section{Experimental Evaluation and Results} \label{Qualitative Analysis and Comparison}
  \subsection{Evaluation}
  \label{sec:evaluation}
  For our comparative analysis, we have used four evaluation metrics:
  
  \[Accuracy=\frac{TP+TN}{TP+TN+FP+FN}\]
  \[Precision=\frac{TP}{TP+FP}\]
  \[Recall=\frac{TP}{TP+FN}\]
  \[F1\;Score=2\times \frac{Precision\times Recall}{Precision+Recall}\]
  True Positives ($TP$) is the number of correctly identified malicious samples. True Negatives ($TN$) is the number of correctly identified benign samples. False Positives ($FP$) is the number of samples that were benign but identified as malicious. False Negatives ($FN$) are the samples that were malicious but not identified correctly by the model. 
  \\\indent Accuracy is a measure of correct classification. Precision is a measure of accurate positive predictions over the total amount of positive predictions. Precision is important because if the precision is low, then the model is predicting many benign samples to be infected. In the case of  cloud data centers, this can hurt the availability of many services if their samples are being incorrectly classified as malign. Recall is a measure of true positive over total actual positive. This metric is important because it reveals how often infected samples get through the model without detection. Recall is useful when the cost of a false negative is high, such is the case with identifying malware. The F1 score is used whenever there needs to be a balance between Precision and Recall and there is a large imbalance in the dataset.
   \begin{table}[t]
      \centering
      \caption{Results for Different Evaluation Metrics}
      \setlength{\tabcolsep}{.85em} 
      \renewcommand{\arraystretch}{1.2}
      \begin{tabular}{c c c c c c}
          \hline\noalign{\smallskip}
          Model & Accuracy & Precision & Recall & F1 & Detection Time (ms) \\
          \noalign{\smallskip}
          \hline
          \noalign{\smallskip}
          LeNet-5 & 89.2 & 94.7 & 80.9 & 87.2 & 54\\
          ResNet-50 & 88.4 & 86.0 & 88.9 & 87.4 & 96\\
          ResNet-101 & 86.6 & 82.3 & 89.7 & 85.9 & 130\\
          ResNet-152 & 89.5 & 89.0 & 87.8 & 88.4 & 165\\
          DenseNet-121 & 92.9 & 100 & 84.6 & 91.5 & 164\\
          DenseNet-169 & 92.8 & 99.7 & 84.4 & 91.4 & 209\\
          DenseNet-201 & 92.8 & 99.5 & 84.6 & 91.5 & 249\\
          \hline
      \end{tabular}
      \label{tab:results_table}
  \end{table}
  \subsection{Experiment Results}
  \indent \autoref{tab:results_table} shows the results of each the CNN models considered in this research. While each model was tested over the course of 100 epochs, these numbers were taken from the model when it scored the highest on the validation data set. This means that these are the best case scenario for each model. If these models were deployed in a cloud environment, they would be trained up to the point at which they generate the best results. This point could be different for every model so it is important to pick out the best performing models and not compare models based on something arbitrary such as after $n$ epochs.
  \\\indent The dataset used consisted of 113 data collection experiments which were split up into the following: training dataset (60\%), validation data (20\%), and testing data (20\%). The training dataset was shuffled but the validation and testing dataset were not. DenseNets reached the highest accuracy at almost 93\% and precision at ~100\%. DenseNets also had the highest F1 scores at ~91.5\%. ResNet-101 had the best recall score at 89.7\%. 
  \begin{figure}[h]
      \centering
      \makebox[0pt]{\includegraphics[scale=0.4]{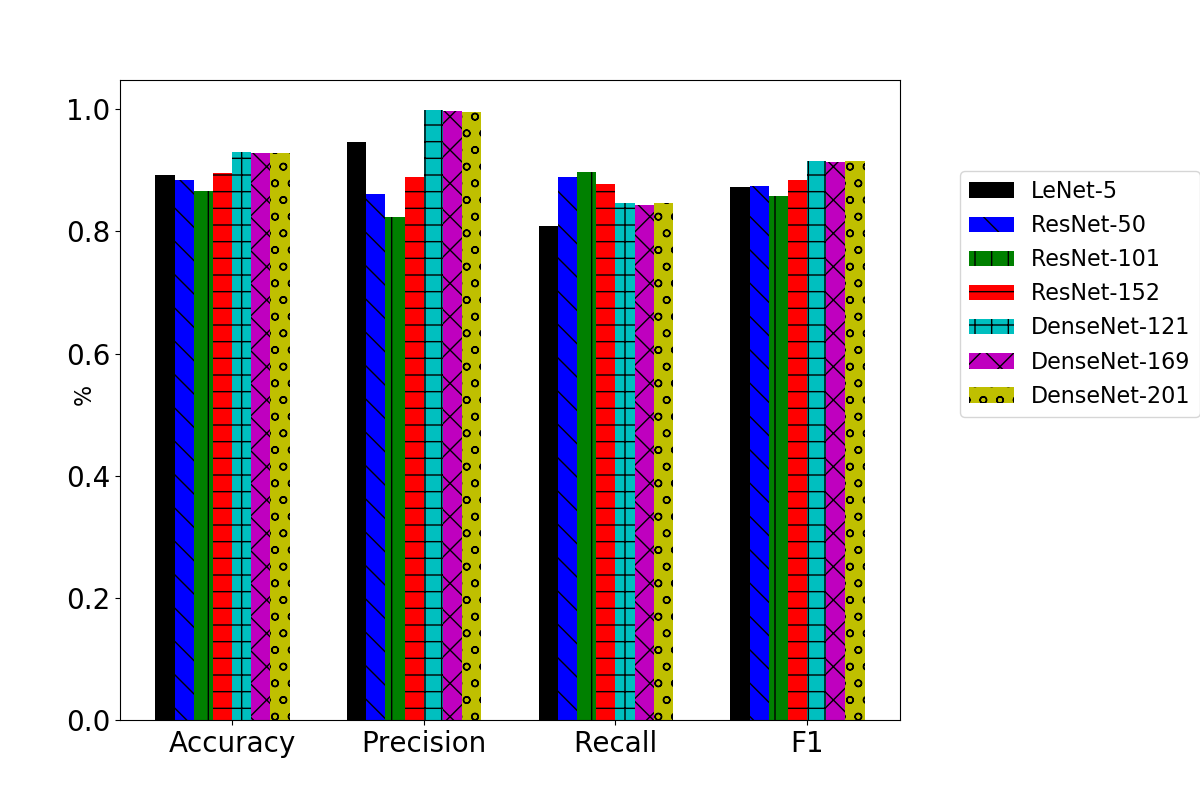}}
      \caption{Metrics Comparison for used CNN Models}
      \label{fig:metrics_bar_graph}
  \end{figure}
  
    \begin{figure}[t]
      \centering
      \includegraphics[scale=0.5]{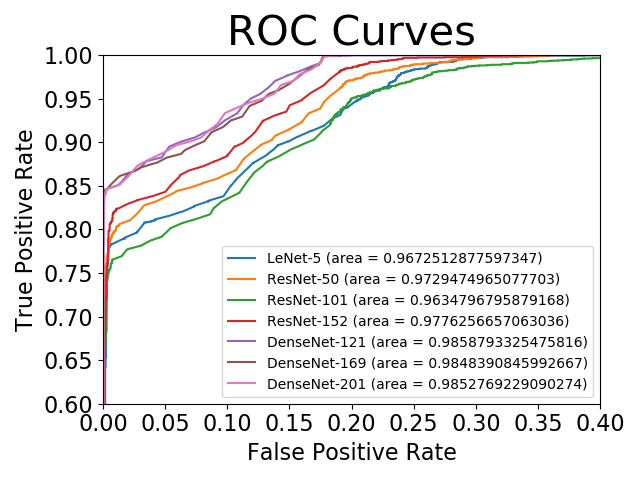}
      \caption{ROC Curves}
      \label{fig:ROC}
  \end{figure}
  \section{Comparative Analysis and Discussion} \label{Discussion}
  As stated in section~\ref{sec:evaluation}, the comparative analysis is performed using four metrics. We discuss each of the metrics along with the ROC curves. Additionally, we discuss the detection time of the models. Finally, we provide an overall analysis discussion and take away which sheds the light on the importance of finding the balance in choosing right models based on the use case and intention. Results for all performance metrics are show in \autoref{fig:metrics_bar_graph}.
  \subsection{Performance Analysis}
  \textbf{Accuracy.} The base model LeNet-5 reaches an accuracy of ~89\%. This is expected as it is a shallow model, thus it lacks the ability to capture enough features. The DenseNet-121 model has the highest accuracy of ~93\%, with a very negligible difference compared to DenseNet-169 and DenseNet-201. This indicates that the adding more layers did not increase the accuracy. One reason might lie in the fact that our dataset is limited (i.e., 40k samples) and deeper networks need more data.
  
  ResNet-152 has a slightly better accuracy than LeNet-5. Considering the substantially longer training time for ResNet-152, such slight accuracy increase from LeNet-5 might not be worthwhile in some cases. Note that ResNet might perform better considering other metrics and, in turn, might work in different scenarios. ResNet-50 and ResNet-101 have the lowest accuracy.
  
  The DenseNets performed better than the other models likely due to the feature reuse property of the dense blocks. Also, DenseNet models are more feature efficient than the other models.

  \textbf{Precision}. The DenseNet models highly outperformed the other models in precision. DenseNet-121 achieved a precision of 100\%, meaning that every sample classified as infected was indeed infected. DenseNet-169 also achieved a high precision of 99.7\% followed by DenseNet-201 with a precision of 99.5\%.
  
  The ResNet models have noticeable lower precision than all the other models, indicating that they are incorrectly classifying benign samples as malicious. LeNet-5 achieved a high precision score so it would be more appealing than the ResNet models when some false positives can be tolerated. The high precision achieved by all the DenseNet models indicates that they correctly identified the benign samples more often and were less likely to classify samples as malicious unless they had a high confidence.

  \textbf{Recall}. Recall is the only metric where ResNets performed better than the other models. All three ResNet models were close but ResNet-101 was the best. The DenseNet models performed worse than the ResNet models but LeNet-5 performed the worst by far. Since recall is a measure of many infected samples where missed by the models, ResNets seem to be effective at identifying most infected samples. LeNet-5's low recall score suggests that the model is weak at identifying less obvious malicious samples. This would be a large problem in datacenters where the samples taken should represent an unbalanced dataset. There should be an overwhelming amount of benign samples before machines are infected and malicious samples begin to show up and a low recall scoring model would be less reliable in predicting the malware as soon as it appears. The higher recall scores demonstrated by the ResNet models are caused by the model being more sensitive and classifying more samples as malicious. This means that the model predicted a sample was malicious more often and was better at identifying those malware who were not as "obvious" in the performance metrics.
  
  \textbf{F1 Score}. F1 Score is about the balance of precision and recall. In that regard, the DenseNet models scored the highest, which indicates that they have the best balance between identifying only malicious samples and identifying most of infected samples overall.
  
  \textbf{ROC Curves}. The receiver operating characteristic (ROC) analysis \cite{metz2006receiver} is used for comparing models at different thresholds. Our ROC curves are shown in \autoref{fig:ROC}. The ROC characteristic measures a models ability to distinguish between classes so in our experiments, it measures the models' abilities to detect malware. If the ROC curve for a model is close to representing a $f(x)=x$ line, then the model has little ability to differentiate between classes. A common way to analyze the ROC curve is to measure the area-under-curve (AUC) value. When the AUC is higher, then the model is accurately predicting benign samples as benign and malicious samples as malicious. The best performing models were the DenseNet models due to their high precision scores which involve both TP and FP values.
  
  \subsection{Cost Analysis}
  \textbf{Training Time}. \autoref{tab:training_time} shows the training time needed to reach the respective accuracies for the models. LeNet-5 trained ten times faster than the next fastest model making it viable as a model to quickly process large volumes of data. DenseNet-201 and DenseNet-169 took much longer to train than DenseNet-121 while reaching similar accuracy making them less desirable.
  

    \begin{figure}[t]
      \centering
      \includegraphics[scale=0.4]{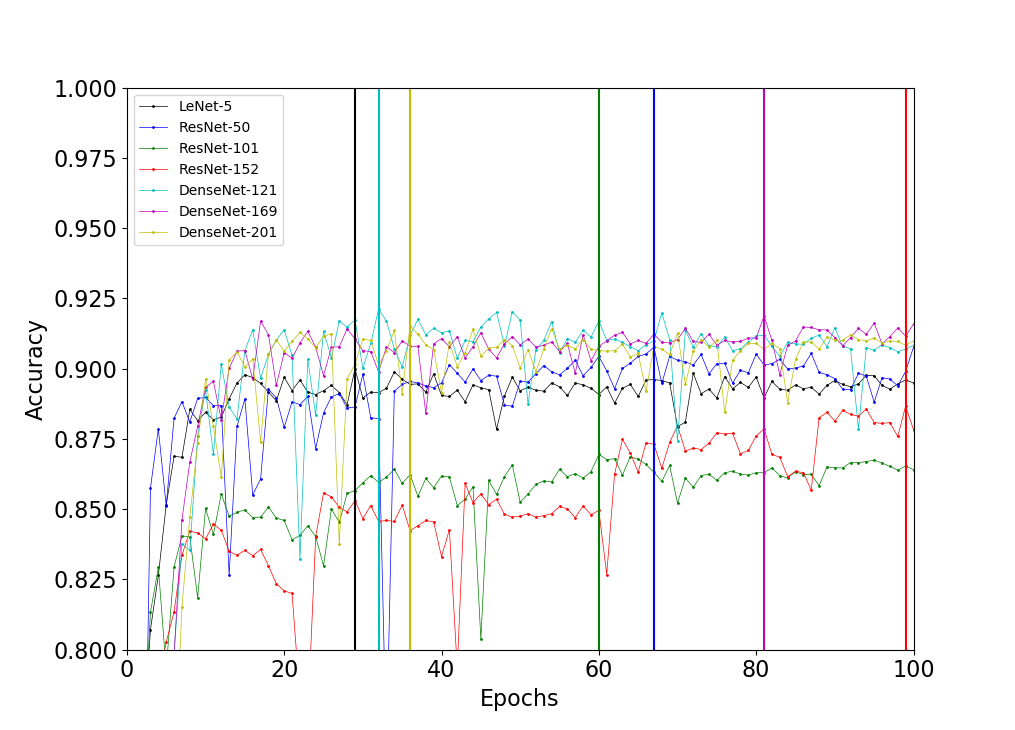}
      \caption{Highest Validation Accuracies Achieved}
      \label{fig:accuracies}
    \end{figure}
    
    \begin{table}[ht]
    \centering
    \caption{Time to Reach Highest Accuracy}
    \setlength{\tabcolsep}{.85em} 
    \renewcommand{\arraystretch}{1.2}
    \begin{tabular}{c c c c c}
      \hline\noalign{\smallskip}
      Model & Validation Accuracy & Epoch Reached & Time Elapsed (s)\\
      \noalign{\smallskip}
      \hline
      \noalign{\smallskip}
      LeNet-5 & 89.9 & 29 & 170\\
      ResNet-50 & 90.7 & 67 & 1815\\
      ResNet-101 & 87.0 & 60 & 2940\\
      ResNet-152 & 88.7 & 99 & 7029\\
      DenseNet-121 & 92.1 & 32 & 1683\\
      DenseNet-169 & 91.9 & 81 & 5848\\
      DenseNet-201 & 91.5 & 36 & 3060\\
      \hline
      \label{tab:training_time}
    \end{tabular}
    \end{table}

  \textbf{Detection Time}. Detection time is used to show how long in milliseconds each model took to produce a prediction for any given sample. The results are unsurprising, more layers in a model cause it to take longer to feed the input through the model. This is important to include, however, because samples in a data center may be getting collected faster than a given model process a prediction. The detection time differences may also indicate that some models may not be suited for lower specification hardware. Since the detection time is dependent on how quickly the model can process the input, increasing the input size or the volume of inputs could prevent some models from scaling with large data center operations. In these cases, the models with lower detection times may be preferable.
  
  \begin{figure}[h]
      \centering
      \makebox[0pt]{\includegraphics[scale=0.45]{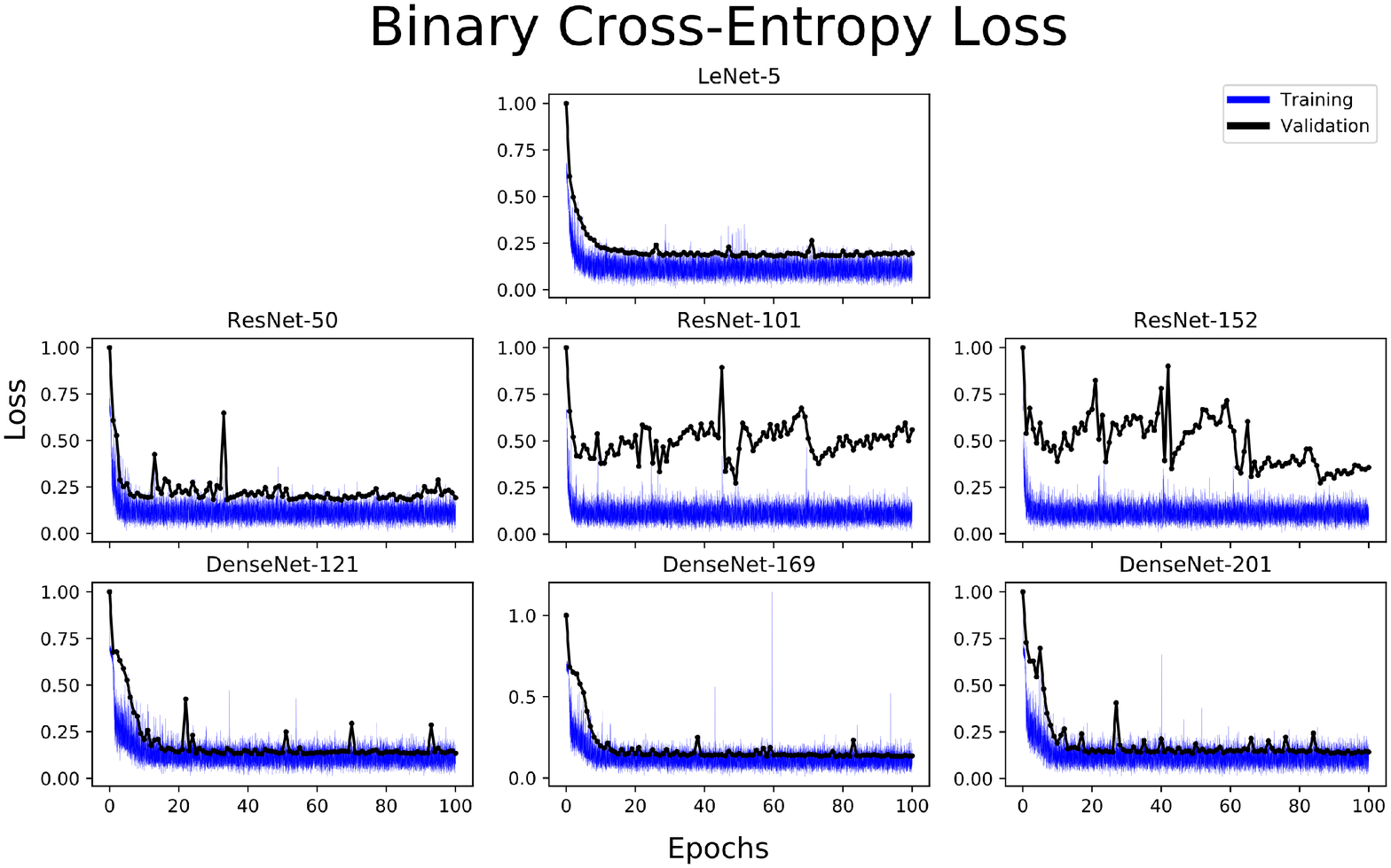}}
      \caption{Training and Validation Loss for used CNN Models}
      \label{fig:loss}
  \end{figure}
  
  \subsection{Overall Analysis}
  Overall, the DenseNet models were the most accurate models with the best balance between precision and recall. The low scores in recall though might be an issue for our use case where allowing malicious samples to slip through could be disastrous. It is also worth noting that while most of the models had validation accuracies that converged to a value, ResNet-101 and ResNet-152 had large fluctuations and never seemed to settle in to a value. This can be seen in \autoref{fig:loss}. If ResNet-101 and ResNet-152 converged to some values, then they may have had better scores. With the inclusion of detection time, assuming the volume of input does not overwhelm the model predicting capabilities, then the DenseNet models would be preferable due to their high accuracy and near perfect precision.
  \\\indent \autoref{fig:accuracies} and \autoref{tab:training_time} show the points in the training where the models reached their highest validation accuracy. The time elapsed column shows the total time needed to reach the epoch where those highest accuracy numbers were achieved. For examples, DenseNet-121 reached its highest accuracy after 32 training epochs and it took 1683 seconds. This shows that DenseNet-121 could be trained for less time than DenseNet-169 or DenseNet-201 and attain better accuracy.

  \section{Limitations and Challenges} \label{Limitations}
  Although, our results provide good understanding of which CNN model works best in what kind of scenario, there are some limitations we would like to highlight based on our experience. The most important limitation of using CNN on the type of data we used is that it fails to capture a time correlation in the data set. When detecting malware in an already running virtual machine, it is important for a model to have some knowledge about existing samples and the behavior of the machine over time. One such scenario is when a machine begins to experience more traffic and due to some constraint on scaling, the samples generated from that machine begin to resemble some malicious samples. In this case, if the model does not learn that process metrics can be scaled according to valid demands on the machine, the false positive rate might increase. Another scenario is when the model detects an infected sample, but the malware immediately becomes dormant as to hide itself. If the model does not take into account the previous sample when the malware was detected, it may increase false negatives where the model doesn't detect a malware even if it is hidden.
  \\\indent These limitations can be mitigated by using Recurrent Neural Networks (RNN). RNN's are comprised of cells which have a memory mechanism and can learn relationships among data with respect to time. These RNN models are used to process sequences of data such as audio or text. Our brief introduction to RNN models suggest that they can be used to solve some of the issues discussed above by lowering false positives and false negatives in certain scenarios.
  
 Another limitation of this paper is the number of malware samples used. We used roughly 120 malware samples, however, we believe with more samples CNN models could have performed better. The deeper networks such as DenseNet-201 and ResNet-152 may perform better on malware that affect the system very little, and the complexity of those networks may be trained on those samples better than a shallower model. By increasing the amount of malware available, the models also gain a broader data set that could be used to better generalize their predictive power. Also once malware is injected, there are no guarantees that the malware is exhibiting malicious behavior at any given time without knowing what code was being executed at that same moment the sample was recorded. This can lead to a problem where samples are mislabeled as malicious or benign. This problem was addressed in \cite{abdelsalam2018malware}, but without writing custom malware that will beacon when malicious activity begins and ends, it is unlikely that all samples will be labeled properly.
  
  
  \section{Conclusion and Future Work} \label{Conclusion}
  In this paper, we analyzed seven different convolutional neural network models to determine which one is better suited for malware detection in cloud IaaS. Our analysis shows that LeNet-5 model is quick but sacrifices accuracy. The model is still useful as it attains a 90\% accuracy and can be used in situations where a quick prediction is needed but incorrectness is not too costly. It can also be used when early predictions can be made with LeNet-5 which can be rechecked with more complex models. Also, our analysis suggest that while the residual networks performed well averaging ~86 accuracy, the DenseNet models performed the best at 93\% accuracy. The ResNet models have higher recall scores indicating that they are more suited for cases where not identifying the malware posing a great security risk. DenseNet models have higher accuracy and precision which indicates they are less likely to generate false positives which are useful in IaaS environments where service availability is extremely important.
  For future work, we plan to examine more malware samples including Windows malware as well as examine other architectures such as Hadoop and containers. We also plan to analyze and propose new deep learning techniques by infecting multiple VMs to replicate more sophisticated attack scenarios. 
  
  
  
  
  \bibliographystyle{unsrt}
  \bibliography{bibliography}

\end{document}